# From Sicilian mafia to Chinese "scam villages" [*]

Jeff Yan

Linköping University, Sweden
jeff.yan@liu.se

**Abstract.** Inspired by Gambetta's theory on the origins of the mafia in Sicily, we report a geo-concentrating phenomenon of scams in China, and propose a novel economic explanation. Our analysis has some policy implications.

## 1    Introduction

Various scams and cybercrime have hit China hard in the recent decade. Notably, seven regions, scattered across the country, were notorious for being scammer hyperactive, and in 2015 the national law enforcement publically announced these regions as high-priority targets in their crackdown of scam and cybercrime.

A look at the notorious scam regions reveals the following. First, they spread across seven provinces including Fujian, Guangdong, Guangxi, Hebei, Hainan, Hunan, and Jiangxi, and the notorious regions are not geographically adjacent. A map will be included here for readers who are not familiar with the geography of China. Second, it is not that everyone in each notorious region was a scammer, but a village or a few in the region was scammer rich, which we call 'scam villages'. Third, the density of scammers in the 'scam villages' was high. For example, reportedly, 50% household of a couple of villages participated in scams, and hundreds of scammers were arrested there [10]. Fourth, the scam economy appeared to be a major trade for each of the scam villages. The scammers made a good living by preying on victims across the country via Internet or telecomm services.

---

Of particular interest, we observe a geo-concentration phenomenon of scams. Namely, a number of villages made a living as scammers; while abundant scammers were inside the villages, the number of scammers quickly diminished outside the village borders. We are not naively suggesting that no scammers exist outside of these villages, but the density of scammers inside the villages and outside differed phenomenally.

Criminology literature can explain to an extent this geo-concentration phenomenon, e.g. the differential association theory [8], which provides a subcultural perspective to explain the wide spreading of criminal behavior; and techniques of neutralization [9], which theorise a series of psychological methods that criminals employ to switch off their inner moral controls.

These theories might suggest that the wide spreading of the scams in a village was the consequence of human interactions. Individuals learned the values, attitudes and techniques for criminal behavior, through interacting with others in the village. Importantly, they learned from others not only the knowledge and skills required for executing each scam, but also neutralization techniques needed to keep their own inner peace. Individuals adopted the neutralizing techniques to 'suspend' negative values attached to a certain behaviour and normalize it. This way, unethical and illicit behaviours were legitimized and justified by the individuals, and by the group as a whole.

Here, we attempt to develop an alternative explanation of the phenomenon, from the perspective of economics rather than criminology. Our focus is not about explaining the formation procedure of the scam villages, but the economic force and rationale behind their formation and existence.

## 2 Related work

Gambetta [5] developed an interesting theory explaining the origins of mafia in Sicily. In his theory, the mafia emergence was a perverse response to a rapid transition to the market economy in the early 19th century, when a demand suddenly surged for the protection of the newly granted property and property rights. While security provided by the state was scarce and banditry widespread, a supply of disbanded soldiers from the Bourbon army and unemployed guards who used to worked for the feudal lords, emerging just about the same time, met the rising demand of protection; they started to offer private protection as a business, giving rise to the Sicilian mafia.

Bandiera [2] further developed Gambetta's property rights theory of mafia emergence. Her analyses suggested that it was optimal for every landowner in Sicily to voluntarily purchase protection from the mafia, even if this would lead to an inferior equilibrium for the landowning class as a whole, and that all things equal, mafia would profit more and develop better where land was more fragmented.

Farrell [4] noticed that "'Crime has a tendency to concentrate in time, space, and other dimensions along which it occurs', and surveyed concepts and terms reflecting various crime concentration theories, including repeat offending, repeat and near repeat victimization, crime hotspots, hot products, hot dots, hot places, hot targets, super-targets, risky facilities, risky routes, and crime sprees and spates. While these are interesting and useful, they are orthogonal to what we intend to achieve in this paper.

Levitt and Venkatesh [7] analysed gang economics with a rare dataset obtained from a drug-selling street gang in the USA. They found that compensation within the gang was highly skewed, and they argued that the prospect of future riches, instead of current wages, was the primary economic motivation for most gangsters. They also suggested that economic factors alone were unlikely to adequately explain individual participation in the gang.

A news article by Wang [10] looked at one of the scam regions in China, but it did not formulise the geo-concentration phenomenon or provide an economic explanation.

Gambetta's study of mafia emergence [5], together with Wang's piece [10], was a major inspiration for us. We don't apply Gambetta's theory in our analysis, but we similarly aim to explain a crime phenomenon from an economic perspective. On the other hand, Wang provided a few key clues to the jigsaw puzzle that our analysis will piece together.

## 3 An economic explanation

### 3.1 The phenomenon: a closer look

Some patterns emerge when we pay a closer look at the phenomenon of 'scam villages'.

First, it occurred only in rural areas, but never in urban communities.

Second, each of the seven regions gained notoriety particularly for a distinctive scam. There appeared to be a clear division of labor, and each of the regions specialised in a particular scam.

Third, these seven representative scams have a clear pattern. 1) To pull off each of the scams, collaborative team efforts were required. Just like the 'police investigation scam' [13] we analysed before, a number of scammers operated in a team, each playing a different role, e.g. extracting victims' personally identifiable information (PII); exploiting technical mechanisms to reach potential victims, and when necessary, to spoof legitimate phones and websites; mind-manipulating victims via impersonation, deception and other psychological tricks; withdrawing money which victims transferred via banks. 2) A lot of efforts were required of the scammers before a victim became hooked to transfer money to them, and for the scammers, the economic yield was near the end of each case. Therefore, both collaboration and trust were required within a team. 3) Not all people targeted by the scammers eventually became a victim that was deceived to lose money. Some targets might narrowly escape from falling into victims near the last minute, and others might realise the scam earlier. It could take time and effort to catch a real victim. 4) Compared to starting up other business, each of the scams was relatively low-cost and lucrative. And it was duplicable. 5) The scams could not be executed in a completely automatic way or without human involvement. Oral communications with victims were needed. Although the scammers could more or less follow a script in their interaction with victims, spontaneous reactions were necessary.

Fourth, not merely a single team, but many teams, were operating the same scam in each of the concerned villages. However, the other villages, geographically adjacent or nearby, might be just normal and they did not home a large number of scammers at all.

The particular geo-concentration feature of the phenomenon that we attempt to explain can be clarified as follows. The practice of a particular scam was propagated inside a village significantly enough to an extent that the density of scammers in the village was high, but the propagation diminished outside the physical boundary of the villages.

Why so?

### 3.2 An analysis

While everyone is selfish (as per the fundamental assumption of economics), savvy scammers who first recognised the profitability of a scam and started a team operation might be willing to share their trade with others due to economic incentives, e.g. if they could benefit from this sharing behavior, and in particular if the (economic) benefit of sharing could significantly outperform that of keeping the trade within their single team.

Under what conditions would the savvy scammers share their trade with others? Let us look into some plausible specifics.

First, the market of the scam economy was large enough. There should be many more victims in the wild than their own team could cope with single-handedly. If the team's income was reduced as a consequence of sharing the trade with others, they would rather keep it to themselves to rake competitive or monopolistic advantages. Or, if the market scale was merely enough to sustain a single team or even smaller, they would not share at all. On the other hand, new teams started with the savvy scammer's help could make a profit for themselves, too.

Clearly, China means an opportunity for a large scam economy to form, due to its vast population of 1.3 billion people, the same language used among them, the same culture shared, and a fast-growing economy in the country.

Second, the range of sharing was not unlimited. If everybody became able to do the trade and did it, the average income of the scammers would decrease due to increased competition. Some barriers should be in place to keep the sharing range within a limit, so that the scammers could maintain a competitive edge in the scam economy.

Third, the economic return of sharing the trade with others was enforceable in practice. The difficulty of enforcing economic returns included the following. 1) This context was all about illegal activities, where people involved would not be naive to rely on law enforcement or the like to enforce their expected return. 2) A key property for a scam team was their knowledge about the scam, i.e. know-hows and sort of 'trade secrets' – not as confidential in the sense of Coca-Cola recipe, but their confidentiality matters[†], since the value of these knowledge diminishes quickly if too many scammers have learned it. However, knowledge sharing is physically non-revocable. 3) Transaction costs can be high for enforcing the rights of trade secrets and business know-hows, even in a legitimate world where legal means and facilities are readily available. These costs could rise to a prohibitively high in an illegitimate world. 4) To enforce the property rights[‡] involved, trust and control was essential between the sharing partners.

Last, if a reasonable mechanism was in place to allow controlled sharing with enforceable return, the savvy scammers could exploit it to maximise self-gain.

An option, readily available to the scammers, was ties of blood and patriarchal clan (宗族). Loosely speaking, a patriarchal clan was an extended kinship group of people with the same surname and lineage (or being genealogically related by marriage), worshiping the same ancestors, and following the Confucian ethical code and customs (e.g. filial piety). The patriarchal clan formed a strong solidarity for members; it protected them against economic adversities, and against outside discrimination or grievance. In the meanwhile, the clan maintained a strong exclusivity to non-members.

---

[†] It is in the scammers' interest to keep their tricks confidential, for another simple reason: if everybody knows the tricks, nobody will fall into victims of the scams.

[‡] "A property right is the exclusive authority to determine how a resource is used", as defined by [1]. Property rights of an economic good or resource include a bundle of rights, including the right to use it, the right to earn income from it, the right to transfer it to others, and the right to enforce property rights.

For hundreds of years, from the imperial times through the Republican period up to the rise of the communist regime in 1949, the patriarchal clan system was the most important institution to understand rural China. Villages were the basic unit of Chinese rural society, and the clan system was the basic institution through which the villages were administrated and run. The villages were often named after the clan which was exclusively or dominantly represented in the villages.

As observed by scholars like Max Weber [12] and John King Fairbank [3], traditional social structure in China had some remarkable features: the central ruling power of the state never reached below the level of county (县), and the rural society was (largely) self-governed by the patriarchal clans in the villages. These clans were 'nourished at the grass-roots level by the principle of filial piety, whose great local strength and gentry leadership make them a match, and even more than a match, for the officials who briefly sojourn among them' [11].

In short, a clan carried the responsibility for administrative, political, economic, educational, policing, defensing, and other functions. It also had the unquestionable authority to lay down the law for its members, and it could exercise power to order civil death or punitive exile (driven out of the village).

The land reform led by Mao Zedong's communist revolution entirely eradicated this patriarchal clan system across the country, by outlawing its organization, eliminating its gentry leadership physically, and forbidding its activities of any sort. Instead, a multi-layer administrative hierarchy was for the first time in the history of China installed below the county level to govern the rural society for the post-1949 era.

Although patriarchal clans were effectively stopped by the political movements, both their organization and activity forbidden, the biological ties of blood and lineage among people were there all the time. Most members of a particular clan often settled in physically nearby, and usually together in a village -- this important pattern of residence remained, too.

Therefore, the savvy scammers could readily make use of the ties of blood and patriarchal clan. This kinship network embodied a social structure, offering both solidarity (to members) and exclusivity (to non-members). And it could work as a natural mechanism of trust and control. The scammers could rely on it to reduce transaction costs of defining, allocating, monitoring and enforcing property rights.

In essence, the geo-concentration phenomenon can be explained in economics terms as follows.

First, this is a phenomenon of controlled sharing, which ensures both a limit of the sharing range and a practical enforceability of economic returns from sharing a trade with others. The kinship social network provided a means of enforcing (intellectual) property rights such as scam know-hows and trade secrets between the scammers.

The savvy scammers were better-off by sharing their trade than keeping the trade secret to themselves; their reward from sharing could be multi-fold, including economic gain (either directly from the increased income or from the reduced expenditure in helping with relatives), or in the form of reputation, respect or social status in their social network.

This sharing practice stopped where the enforceability of its economic returns became week, e.g. typically along the physical border of the villages. The other villages, adjacent or nearby, usually belonged to a different kinship social network.

The savvy scammers could maximise self-gains by 1) scaling up productivity and profitability through sharing the trade, 2) controlling the extent of sharing, by limiting the knowledge propagation via a trusted kinship social network only, to prevent the economic value of their knowledge diminishing from a nearly monopoly to an unacceptable low, and 3) maintaining a practical enforceability of expected economic return from sharing the trade with trusted partners.

### 3.3 Evidence

We offer some evidence to support our analysis.

The propagation of knowledge and skills of a particular scam across the kinship social network increased the welfare of not only individuals and families who received the knowledge and skills, but also those givers. Inevitably, this propagation process had done more than activating the kinship ties, but reinforced them. Therefore, if the resurgence of patriarchal clans is observed in the scam villages, that will support our theory. Specifically, if activities of the clan nature in the concerned village reached an intensity level that was higher than the average, this observable fact can corroborate our theory as a reliable evidence.

Indeed a journalist suggested in [10] that the patriarchal-clan organizations did emerge, and substantial activities of that nature were observed, in the few scam villages which the author paid attention to. An official publication of the national law enforcement [6] even suggested that the clan activities became alarmingly intense in the villages.

Nowadays the official authorities in some regions have started to turn a blind eye to certain activities of the patriarchal clan nature, those considered to pose less of a threat to the government, such as compiling genealogical records (which are an important document for each clan), rebuilding ancestral halls (the temples) and organising worship rituals. But the scam villages went way beyond these. Councils and executive committees were set up within a clan, and elected leadership emerged. These committees were independent from and in parallel to the official administration authority, and the former often took effective charge of the villages. Organised resistance was staged to mitigate the crackdowns led by the law enforcement. For some time, the clans had made it difficult or even impossible for the police to get inside the villages making arrests [6]. Clearly, these activities were beyond the red line that the government could tolerate with.

It was also observed that some key scammers served in the leadership committees of their clan. Their rise in social status within the clan was evidently a non-monetary reward from sharing their scam trade with fellow villagers.

As additional evidence to support our theory, it was documented in [10] that less competent teams were helped by more sophisticated ones to collaboratively work on challenging but potentially lucrative targets. And they would share any yield squeezed successfully from the targets. This sort of collaboration makes a lot of economic sense, when false positives and true positives are taken into consideration.

Not all people targeted by scammers will yield anything. If we say false positives are targets that are attacked but yield nothing, then true positives are targets successfully attacked. A key issue in the scammers' business model is to control the ratio and the relative cost of true and false positives. The cooperation of less competent teams and more sophisticated ones could effectively improve the ratio of true and false positives, which was win-win for both teams.

We would not be surprised if some scammers helped new teams to start up so that they could benefit financially in a form of shareholding in the new teams' scam business, i.e. a certain percent of profit raked by the latter was paid to the former as a dividend of sharing their trade. Similarly, we would not be surprised if a team procured sick replacements from another team. However, no public resources confirm or rebuke these predictions, yet.

Several facts explain why this geo-concentration phenomenon occurred only in rural areas, but not in cities. First, it is not that people in urban communities ignore the ties of blood and patriarchal clans, but this kinship bond is significantly weaker in the city than in the countryside. Second, the demographic differs. In the cities, it is rare for people from the same patriarchal clan to settle in physically nearby, and rare for them to settle in at an area as isolated as a village. Third, within the city wall, it has been governed by the central power of the state, and the patriarchal clans have never dominated there, even in the imperial times.

### 3.4 Why didn't 'scam villages' form elsewhere?

It is because the formation of this phenomenon requires a combination of conditions, including: 1) scammers who are smart enough to figure out that scalability trumps 'keeping the trade to myself', 2) a mechanism in place that not only controls the range of sharing, but also effectively enforces the property rights involved, 3) law enforcement that are uninformed of, or unprepared for modern scams, or being corrupt or corruptible, and 4) an environment that is lack of other opportunities for making a good living.

# 4 Conclusions

The 'scam villages' represent a form of economic enterprise. It was built upon an alternative system of property rights which the scammers figured out and adopted to expand their scam business, when legally enforceable contracts or property rights were impossible, due to the illicit nature of their operation.

The ties of blood and patriarchal clans, deeply rooted in Chinese rural society for centuries, provided a mechanism of trust and control. The scammers exploited this mechanism to not only propagate and share their scam knowledge and skills across the kinship social network, but also prevent further sharing and propagation.

In the meanwhile, the ties facilitated a mechanism for reducing the transaction costs of delineating, monitoring and enforcing property rights from a prohibitively high, making it viable to practically enforce the rights among the scammers on the same kindship social network.

The property rights concerned in this context were way beyond intellectual property rights of scam 'trade secrets' and know-hows that started a team up and running, but could include rights to the proceeds squeezed from a victim on which multiple teams worked collaboratively to achieve a success, rights to the proceeds contributed by sick-replacement labor, and so on.

In this 'scam village' form of economic enterprise, many insiders in the village were able to join the business to share the economic benefit allowed by the sufficiently large scam economy, whereas outsiders were excluded. Apparently this arrangement protected the insiders' interest and prevented potential competition from the outsiders, simultaneously.

We have observed some evidence that supports our analysis. Therefore, our economic explanation is more than merely hypothetical, but partially empirical.

As the renowned economist Armen A. Alchian nicely put, "The definition, allocation, and protection of property rights comprise one of the most complex and difficult sets of issues that any society has to resolve, but one that must be resolved in some fashion" [1]. Apparently, people in the scam villages had figured out their own way.

Our analysis has some policy implications. The scam villages are a form of crime where some people have a bigger role or reward, than the others. The law enforcement should identify the few key criminals and hit them hard with a first priority.

Arguably, the formation of each scam village also created a network effect for itself. The scammers in the village competed with each other, and they also collaborated in one way or another. As more people joined this network, the strength of the network kept increasing – the scammers improved their skills on a particular form of scam every day, significantly increasing the competitive edge of them as a whole in the scam economy. This continuous self-reinforcement process might explain why people in each of the seven notorious regions largely specialised in a particular scam. Staying competitive – it was the market force that was working. It was a specialised division of labor.

This might also explain a paradox: as more and more people in the village joined the scam economy, the competition between themselves increased, very likely leading to a decreased average income for everyone involved; but why more and more insiders were allowed to join force afresh? Our explanation is the following. The increased competence of the village as a whole in a particular scam allowed a relatively large number of insiders to join the business by outperforming the competence of outsiders to reduce their participation in the same scam.

In a sense, some scammers stayed close, simply because they were born close. Staying close, they also became stronger.

**Acknowledgement**

I thank Robert Axelrod, Ross Anderson and Dah Ming Chiu for valuable comments. This work was supported in part by EPSRC grant "The Deterrence of Deception in Socio-Technical Systems" (EP/K033476/1) in the UK, and by the Knut and Alice Wallenberg Foundation in Sweden.